\documentstyle[12pt]{article}
\title{{\hfill{\large IC/93/33}} \\
{\hfill{\large hep-th/9302100}} \\
\  \\
 STATIC QUARK POTENTIAL FROM THE POLYAKOV SUM OVER SURFACES}
\author{Zbigniew Jask\'olski\thanks{E-mail: jaskolsk@ictp.trieste.it.
Permanent address: Institute
 of Theoretical Physics, Wroc{\l}aw University, pl. Maxa Borna 9,
50-204 Wroc{\l}aw, Poland}
 \and
 Krzysztof A. Meissner\thanks{E-mail: meissner@ictp.trieste.it.
Permanent address: Institute of Theoretical
Physics, Warsaw University, Ho\.za 69, 00-681 Warszawa, Poland } \\
International Centre for Theoretical Physics,
 Trieste, Italy}
\date{February 17, 1993}
\begin{document}
\titlepage
\maketitle
\abstract{ Using the Polyakov string ansatz for the rectangular
Wilson loop we calculate the static potential in the semiclassical
approximation. Our results lead to a well defined sum over surfaces
in the range $1<d<25$.}
\newpage
\pagenumbering{arabic}

\def \xinfty{\displaystyle_{x\to \infty}^{_{_{--\longrightarrow}}}}
\def \xzero{\displaystyle_{x\to 0}^{_{_{-\longrightarrow}}}}
\def \rinfty{\displaystyle_{\rho\to \infty}^{_{_{--\longrightarrow}}}}
\def \rzero{\displaystyle_{\rho\to 0}^{_{_{-\longrightarrow}}}}
\def \Tinfty{\displaystyle_{T\to \infty}^{_{_{--\longrightarrow}}}}
\def \Rinfty{\displaystyle_{R\to \infty}^{_{_{--\longrightarrow}}}}
\def \RRc{\displaystyle_{R\to R_c}^{_{_{-\longrightarrow}}}}
\def \ap{{2\pi\alpha'}}
\def \apd{{4\pi\alpha'}}

\section{Introduction}

In the present paper we address the question of the static $q\overline{q}$-
potential in the noncritical Polyakov string model \cite{pol} for the Wilson
loop.
The string calculation of this object stems from the old idea \cite{w,ansatz}
according to which the sum over world-sheets with fixed boundary
provides a reasonable low energy approximation for the loop
expectation value in QCD. The problem of the static potential in
the Schild--Eguchi \cite{schild} and in the Nambu--Goto string models was
solved several years ago
\cite{lsw,luscher,dietz,alv,arvis}. Actually it is one of few string
model predictions which can be compared with the Monte Carlo
\cite{montec,montec2}
and the experimental
\cite{exper} QCD data. Since the noncritical Polyakov string
does not fall into the universality class
\cite{luscher} of  models
mentioned above it is desirable to have corresponding results
also in this case.
There is however a more general theoretical context which is in fact
our main motivation. The  long rectangular Wilson loop
- an object with clear physical and geometrical interpretation, yet simple
enough to make explicit calculations possible - is an ideal theoretical
laboratory to investigate  the Dirichlet type boundary conditions
for the Polyakov sum over surface
in noncritical dimensions. It seems that a better understanding of this
problem can shed new light both on the Polyakov string ansatz for the loop
equations and for the quantum physics of the conformal factor itself .

The first seminal calculations of the leading (1--loop) correction to
the loop expectation value in the Schild-Eguchi model was done by
 L\"{u}sher, Symanzik and Weisz \cite{lsw}.
In the particular case of long  rectangular loop they discovered the
Coulomb-like correction to the linear static potential at large distances.
It was subsequently observed by L\"{u}scher \cite{luscher} that this term is
universal for all string models for which the transversal "embedding"
variables are the only relevant degrees of freedom  in the low energy limit.
This is in contrast with the higher order corrections to the effective
"transverse" action which are model dependent \cite{dietz}. An attempt to
probe small distances was done in \cite{alv}, where the leading term in the
$1/d$ expansion was calculated in the Eguchi and in the Nambu-Goto models.
This result exhibited the existence of a critical distance and was further
confirmed by  exact calculations in the light-cone operator quantization
of the critical Nambu-Goto string \cite{arvis}. These string model predictions
were  confronted with Monte Carlo data in a number of papers
\cite{montec}. The general outcome of these analyses is a
confirmation of the effective  string picture at large distances
and in particular the presence of the roughening phenomenon.
The results concerning the numerical value of the coefficient of
the Coulomb-like term  depend on the methods used to disentangle
finite size effects and seem to be insufficient to select a
proper string model \cite{montec2,olesen}.

The covariant functional techniques employed in the present paper are based on
the  discussion of the Polyakov string model for the Wilson loop
due to O.Alvarez. In his pioneering paper
\cite{alvarez} the boundary contributions to
the conformal anomaly along with the explicit forms of integration measures
for Teichm\"{u}ller parameters and the conformal factor were calculated for
bordered surfaces of arbitrary topological type. At that time however,
the only available technique to deal with the Liouville sector was the
saddle point approximation \cite{ft,dop}. A few years ago
the main problem in the quantization
of the Liouville theory - the field dependent functional measure -
was successfully solved by David \cite{d} and
Distler and Kawai
\cite{dk} within the CFT approach. The DDK method was soon rederived by
direct calculations of the functional Jacobian
\cite{mm,dhku}. Since the approach of
\cite{mm,dhku}
does not appeal to the conformal invariance of the combined
ghost - Liouville - matter system it can be firmly used in the range
$1< d < 25$, where the CFT methods break down. This is one of important
points in our presentation.

The second crucial  issue  is the choice of
boundary conditions. For previous discussions of this point we refer to
\cite{alvarez,ft,dop}.
In the present paper we use the boundary conditions uniquely determined by
the consistency conditions of the Faddeev-Popov procedure. They were
first introduced in the context of off-shell critical string
amplitudes
\cite{zj}  and recently analysed in the Liouville gravity
coupled to the conformal matter in \cite{zj2}.
This choice is motivated by
the interpretation of the Polyakov path integral as a sum over
surfaces, which in the continuum formulation
is based on the geometrical content of the Faddeev--Popov
method.

The organization of the paper is as follows. In Section 2 we discussed
the Polyakov sum over surfaces with particular attention paid to
the proper treatment of the boundary conditions. In the case relevant
for calculations of the static potential the formula for the Wilson
loop is derived.
 As a result of our choice of boundary conditions one gets
 a novel approach to the Liouville theory and its interaction
 with the "matter" sector. In particular our formula for the Wilson loop
 does not contain the averaging over boundary reparametrisations
 \cite{cmnp}
 nor the integration over boundary values of conformal factor  \cite{dop}
although the second method may be equivalent to the one presented in this
paper.  The considerations of this section are general and can be applied to
arbitrary contours and model manifolds. In Sect.3 we present all
calculations which can be done exactly in the case of rectangular
loop. For the vanishing and for the strictly positive value of the bulk
cosmological constant $\mu$ one gets two different expressions
containing path integrals over the nonconstant modes of conformal factor.
Being unable to perform these integrals exactly we introduce an
approximation scheme with the expansion parameter $\epsilon =
{\sqrt{2\pi\alpha'} \over R}$. The general features of this scheme are
discussed in Sect.4 in the case $\mu =0$.
Since $\epsilon$ enters the effective action
for $\phi$ only via the interaction term the coefficients in the corresponding
power expansion get additional $\epsilon$-dependence. This also leads to
a nonstandard relation with the loop expansion.
In order to recover the static potential
in the semiclassical approximation we  analyse the limit
of long rectangular loop. In this limit all
functional integrals can be performed. Using the saddle
point approximation of
the resulting integral over the Teichm\"{u}ller parameter we
derive the explicit large $T$
asymptotic from which the static potential can be obtained.
The corresponding
calculations for $\mu > 0$ are presented in Sect.5. On the
semiclassical level the static
potential is the same in both cases.
The comparison with the predictions of string models from L\"{u}scher's
universality class suggests a simple (free) dynamics of the longitudinal
mode.
 Some technical points of our derivation  are explained in
the appendix.
 Finally Sect.6 contains conclusions and some suggestions for
future work.

\section{Boundary conditions}

 We start with a brief description of
the approach of \cite{zj} in the context  of  noncritical string.
For simplicity we consider
a  loop $c \in R^d$ with four corners  and the Polyakov path integral
\begin{equation}
W[c] = \int_{{\cal F}[c]} {\cal D}g {\cal D}x
\left( Vol \mbox{\it Diff} \right)^{-1} \exp - {\scriptstyle
\frac{1}{4\pi \alpha '}} S[g,x]
\label{da}
\end{equation}
over all surfaces of the topology of rectangle bounded by the loop $c$.
The choice of  loop with corners is slightly artificial from the
point of view of the Wilson loop. We have chosen it to illustrate the general
structure of the boundary conditions, first because it is directly
related to the example we will calculate in the following; secondly
it allows to avoid the technicalities related to the $S^1$ action
present in the case of boundary components diffeomorphic to $S^1$
\cite{zj3}.

Let $M$ be a model rectangle with boundary components $ \partial M_i ;
i=1,...,4$;
$ {\cal M}_M $ -- the space of all metrics on $M$, and $ {\cal E}^d_M $ --
the space
of all maps $x:M\rightarrow R^d$. Our aim is to determine the subspace
$ {\cal F}[c] \subset {\cal M}_M \times {\cal E}^d_M $ of "string trajectories"
bounded by the loop $c$ and satisfying
the consistency conditions mentioned in the introduction. Clearly for any
$ (g,x) \in  {\cal F}[c] $
 , $ x_{| \partial M}: \partial M \rightarrow R^d $ must
be a parametrization of $c$. With our ability to perform only gaussian
integrals we need the Dirichlet boundary conditions for $x$ with some choice of
this parametrization. On the other hand in the next step we would like to use
the F-P procedure with respect to the group $\mbox{\it Diff}_M$ which means
that the
result of $x$-integration must be a $\mbox{\it Diff}_M$-invariant functional of
$g$.
These two requirements are in apparent contradiction. Fortunately there is
a simple solution for this puzzle. At the beginning we have tacitly assumed
that
each smooth component $c_i, i=1,...,4$ of $c$ is a point in the space
\begin{equation}
c_i \in \frac{{\cal E}^d_{I_i}}{\mbox{\it Diff}_{I_i}}\;\;\;,
\label{db}
\end{equation}
where $ {\cal E}^d_{I_i}$ is the space of maps
$\widetilde{x}: I_i \rightarrow R^d$
and $I_i , i =1,...,4$ are model intervals. Suppose $c$ has an additional
internal
structure, namely
\begin{equation}
c_i \in \overline{{\cal C}}_i \equiv \frac{ {\cal M}_{I_i} \times
{\cal E}^d_{I_i} }{\mbox{\it Diff}_{I_i} \times R_+} \approx {\cal E}^d_{I_i}
\;\;\;,
\label{dc}
\end{equation}
where ${\cal M}_{I_i}$ is the space of all einbeins on $I_i$ and
$R_+$ stands for the group of constant rescalings of einbeins. (If $c_i =
[(\widetilde{e}_i,\widetilde{x}_i)]$ and
$\widetilde{x} \in {\cal E}^d_{I_i}$ is regular
enough one can regard $c_i$ as a 1-dim submanifold of $R^d$ endowed with
intrinsic metric determined up to a constant multiplicative factor.)
Now let us observe that each "string trajectory"
$(g,x) \in  {\cal F}[c]$ uniquely determines a point
\begin{equation}
[(e_i,x_i)] \in \frac{ {\cal M}_{\partial M_i} \times {\cal E}^d_{\partial M_i}
}
{\mbox{\it Diff}_{\partial M_i} \times R_+} \;\;\;,
\label{dddd}
\end{equation}
where $e_i$ denotes the einbein induced by $g$ on $\partial M$ and
$x_i = x_{|\partial M_i}$. Since the spaces
(\ref{dc}),(\ref{dddd}) are canonically
isomorphic, for any system of $ c_i \in {\cal C}_i$  (satisfying appropriate
compatibility conditions at the ends of $I_i$) the boundary
conditions
\begin{equation}
       [(e_i,x_i)] = c_i \;\;\;, \;\;\; i=1,...,4.
\label{dbc}
\end{equation}
are perfectly well defined and $\mbox{\it Diff}_M$-invariant.
For each metric $g$ these conditions imply a  $g$-dependent Dirichlet boundary
condition for $x$
\cite{zj}. Hence the $x$-integration is gaussian and yields
 $\mbox{\it Diff}_M$-invariant functional $Z[c,g]$ on ${\cal M}_M$
 $$
 Z[c,g] = \left( \det {\cal L}_g \right)^{-\frac{d}{2}} \exp
 -S_{\mbox{\scriptsize  eff}}[c,g]\;\;\;,
 $$
 where the determinant for the scalar Laplace-Beltrami operator is
 calculated for the homogeneous Dirichlet boundary condition.

 Two remarks concerning  the boundary conditions (\ref{dbc})
are in order.
First of all they are not invariant with respect to general Weyl
transformations. This fact leads in the critical string theory to
the gauge dependence of the off-shell amplitudes. In the noncritical
theory it yields nontrivial coupling of the conformal factor to
the matter fields via boundary conditions.
This was first pointed out in
\cite{dop} where a slightly stronger version
of the boundary condition (\ref{dbc}) was proposed.
Secondly the Wilson loop $c$ in
QCD is supposed to be a "nonparametrized contour" i.e. a point in the
space (\ref{db}). Switching over from (\ref{db}) to (\ref{dc}) means
introducing
some preferred parametrization of $c$. When
the loop $c$ is regular enough (locally embedded in $R^d$)
there exists a natural choice - the parametrization in which the
metric induced by the target space metric  is constant.
We will use this parametrization for the rectangular loop
appearing in  calculations of the static potential.
Note that for less regular
loops this simple prescription breaks down.

The next step in the calculation of  (\ref{da}) is to apply the
F-P method to the path integral
\begin{equation}
W[c] = \int_{{\cal M}_M^*} {\cal D}g
\left( Vol \mbox{\it Diff} \right)^{-1} Z[c,g]\;\;\;,
\label{dd}
\end{equation}
where the integration domain ${\cal M}_M^* \subset {\cal M}_M $ is not yet
specified. The problem of functional integration over the space of
metrics on a bordered surface with $\mbox{\it Diff}$ - invariant
integrand has been recently analysed in
\cite{zj2}. It was shown that the consistency
of the F-P procedure uniquely determines a family of admissible integration
domains ${\cal M}_M^{n*} \subset {\cal M}_M $ parametrized by the space
of all normal directions $n$ on the boundary. Moreover the resulting
path integral is independent on $n$ which means that ${\cal M}_M^*$ in
the formula (\ref{dd}) is in fact uniquely determined.

In the case under consideration,
for a fixed normal direction $n$ on $\partial M$ the space ${\cal M}_M^{n*}$
can be described as follows. Let $M^{DD}$ be the double double of $M$
(i.e. the result of subsequent doubling with respect to two pairs of disjoint
boundary components of the rectangle $M$) and $i, j$ - corresponding
involutions of $M^{DD}$ with the invariant direction $n$ on $\partial M$.
Then the space ${\cal M}_M^{n*}$ consists of all metrics on $M$ admitting
a $C^1$-extension to an $i,j$--symmetric metric on $M^{DD}$. Another
description of ${\cal M}_M^{n*}$ can be given in terms of the space
${\cal M}_M^{no}$ of all metrics on $M$ with the normal direction $n_g = n$
and the scalar curvature $R_g = 0$ such that all boundary components are
geodesic and meet orthogonally. One can show that
${\cal M}_M^{n*}$ consists of all metrics of the form $e^{\phi}g_o$
where $g_o \in {\cal M}_M^{no}$ and $\phi$ satisfies the Neumann boundary
condition $n^a{\partial}_a \phi = 0$ on $\partial M$.
The most important consequence of this result is that in every
conformal gauge $\mbox{$R_+ \ni t \rightarrow \widehat{g}_t
\in {\cal M}_M^{n*}$}$, $g = e^{\phi}\widehat{g}_t$
the conformal factor satisfies the Neumann boundary condition
\begin{equation}
n^a {\partial}_a \phi = 0
\label{de}
\end{equation}
on each boundary component.

With the choice of ${\cal M}_M^{n*}$ as an integration domain
in (\ref{dd}) the F-P procedure in a
conformal gauge yields
\begin{eqnarray}
W[c] = \int\limits_0^{\infty} [dt]
\int\limits_{{\cal W}^n}\!\! {\cal D}^{\widehat{g}_t}
\phi
&\times&\exp{\left(- {25-d \over 48\pi} S_L[\widehat{g}_t,\phi]\right) }
 \nonumber \\
&\times& \exp{\left( {7+d \over 32} \sum\limits_{\mbox{\scriptsize corners}}
\phi(z_i) \right)} \label{df}\\
&\times& \exp{\left(- {1 \over 4 \pi \alpha'}
S_{\mbox{\scriptsize eff}}[c,e^{\phi}\widehat{g}_t] \right)}\nonumber
\end{eqnarray}
where $[dt]$ stands for the $\phi$-independent part of the integrand
and
${\cal W}^n$
is the space of all realvalued functions on $M$
satisfying the boundary condition (\ref{de}). For the calculation of the corner
contribution to the conformal anomaly we refer to the original paper
\cite{vw}
where this formula was first derived.
As a consequence of our choice  of integration domain in (\ref{dd})
the Liouville action in (\ref{df})
$$
S_L[g,\varphi] = \int\limits_{M_{h,b}}\!\!\! d^2\!z \sqrt{g} \left(
{\scriptstyle {1 \over 2}}g^{ab} \partial_a \varphi \partial_b \varphi
+ R_g \varphi + \mu e^{\varphi} \right)
$$
does not contain boundary terms. In particular the boundary conditions
(\ref{de}) imply the vanishing boundary cosmological constant. The
action is bounded from below for the bulk cosmological constant
$\mu \geq 0$.

\section{ Rectangular Wilson loop}

In order to make our abstract discussion more concrete let us introduce
the 2-dimensional
\begin{equation}
(\widehat{M}_t,\widehat{g}_t) = ([0,t]\times[0,1],
 \left( \parbox{16pt}{ \scriptsize
1 \makebox[1pt]{} 0 \\
0 \makebox[1pt]{} 1 }
 \right) ) \;\;\;,
 \label{dg}
\end{equation}
and the 1-dimensional
\begin{equation}
\begin{array}{lcll}
(\widehat{I}_i = \partial \widehat{M}_{ti} , \widehat{e}_i ) & = &
([0,t],1) & \;\;\; i =1,3\;\;\;,\\
(\widehat{I}_i = \partial \widehat{M}_{ti} , \widehat{e}_i ) & = &
([0,1],1) & \;\;\; i =2,4\;\;\;,
\end{array}
\label{dgo}
\end{equation}
conformal gauges. The measure $[dt]$
involving determinants of $P^+_{\hat{g}_t}P_{\hat{g}_t}$ and
${\cal L}_{\hat{g}_t}$
can be calculated by standard methods. In the conformal gauge (\ref{dg})
one gets
\cite{cmnp,vw}
$$
[dt] = \eta (t)^{1 - \frac{d}{2}} dt
$$
where
$$
\eta (t) = {\rm e}^{-\frac{\pi t}{12}}\prod_{n=1}^{\infty}\left(
1-{\rm e}^{-2\pi n t}\right)\;\;\;.
$$
Let $(\widetilde{x}_i,1)$ be the representant of
$c_i$ in the gauge (\ref{dgo}). According to \cite{zj}
\begin{equation}
S_{\mbox{\scriptsize eff}}[c,e^{\phi}\widehat{g}_t] =
S[\widehat{g}_t,\widetilde{x}_i[\phi] ] = \int^t_0 dz^0 \int^1_0 dz^1
\left(\left({\partial}_0 x_{cl} \right)^2 + \left({\partial}_1 x_{cl}
\right)^2\right)
\label{dh}
\end{equation}
where $x_{cl}: \widehat{M}_t \rightarrow R^d$ is the solution of the boundary
value problem
\begin{eqnarray}
& & \left( {\partial}^2_0 + {\partial}^2_1 \right) x = O \nonumber\\
& & x_{|{\partial}M_i} = \widetilde{x}_i[\phi] \equiv
\widetilde{x}_i \circ \gamma_i[\phi]\;\;\;,i=1,...,4.
\label{dk}
\end{eqnarray}
The diffeomorphisms $\gamma_i[\phi] : [0,t] \rightarrow [0,t], i=1,3$ and
$\gamma_i[\phi] : [0,1] \rightarrow [0,1], i=2,4$ are uniquely determined
by the equations
\begin{eqnarray}
\frac{d}{dz^0} \gamma_1[\phi](z^0) & \propto & \exp {\scriptstyle \frac{1}{2}}
\phi(z^0,0) \;\;,\nonumber\\
\frac{d}{dz^1} \gamma_2[\phi](z^1) & \propto & \exp {\scriptstyle \frac{1}{2}}
\phi(0,z^1) \;\;,\nonumber\\
\frac{d}{dz^0} \gamma_3[\phi](z^0) & \propto & \exp {\scriptstyle \frac{1}{2}}
\phi(z^0,1) \;\;, \label{di}\\
\frac{d}{dz^1} \gamma_4[\phi](z^1) & \propto & \exp {\scriptstyle \frac{1}{2}}
\phi(t,z^1) \;\;.         \nonumber
\end{eqnarray}
In general the prescription given above yields the functional
$S[\widehat{g}_t,\widetilde{x}_i[\phi] ]$ depending on the boundary
value of $\phi$ in a very  complicated way. The situation
is more tractable for rectangular loops. Let $c_{RT}$ be a loop
of width $R$ and length $T$ given in the 1-dim conformal gauge (\ref{dgo}) by
\begin{equation}
\begin{minipage}{100pt}
\begin{eqnarray*}
\widetilde{x}^0_1(z^0) & = & {\textstyle \frac{T}{t}} z^0 \;\;,\\
\widetilde{x}^0_2(z^1) & = & 0 \;\;,\\
\widetilde{x}^0_3(z^0) & = & {\textstyle \frac{T}{t}} z^0 \;\;,\\
\widetilde{x}^0_4(z^1) & = & T \;\;,
\end{eqnarray*}
\end{minipage}
   \makebox[40pt]{}
\begin{minipage}{100pt}
\begin{eqnarray*}
\widetilde{x}^1_1(z^0) & = & 0 \;\;,\\
\widetilde{x}^1_2(z^1) & = & R z^1 \;\;,\\
\widetilde{x}^1_3(z^0) & = & R \;\;,  \\
\widetilde{x}^1_4(z^1) & = & R z^1 \;\;,
\end{eqnarray*}
\end{minipage}
\label{dj}
\end{equation}
$$
\widetilde{x}^k_i \equiv 0 \makebox[20pt]{} \mbox{for} \makebox[20pt]{}
i=1,\ldots ,4\;; \;k=2,\ldots , d-1\;\;\;.
$$
Solving the boundary value problem (\ref{dk}) and inserting solution into
the Polyakov action one gets
\begin{eqnarray}
S[\widehat{g}_t,\widetilde{x}_i[\phi] ] & = &
\frac{T^2}{t} +
2T^2
\sum\limits_{n > 0}
\frac{ 1}{\pi n \sinh {\displaystyle \frac{\pi n}{t} } }
\left[ ( {f_{1n}}^2 + {f_{3n}}^2 ) \cosh \frac{\pi n}{t} -
2 f_{1n}f_{3n} \right] \label{dl} \\
& + & R^2 t +
2R^2
\sum\limits_{m > 0}
\frac{1}{\pi m  \sinh \pi mt }
\left[ ( {f_{2m}}^2 + {f_{4m}}^2 ) \cosh \pi mt -
2 f_{2m}f_{4m} \right] \;\;\;,   \nonumber
\end{eqnarray}
where
\begin{eqnarray}
f_{1n} & = & \frac{\pi n}{Tt}  \int\limits_0^t  dz^0
\left( {T \over t} \gamma_1[\phi](z^0) - {T \over t} z^0 \right)
\sin \frac{\pi nz^0}{t}  \nonumber \\
& = &  \left( \int\limits_0^t dz^0
\exp{{\scriptstyle {1 \over 2}} \phi(z^0,0)}  \right)^{-1}
\int\limits_0^t                 dz^0
\exp{{\scriptstyle {1 \over 2}} \phi(z^0,0)}
\cos \frac{\pi n z^0 }{t}  \nonumber \\
f_{2m} & = &
\left( \int\limits_0^1 dz^1
\exp{{\scriptstyle {1 \over 2}} \phi(0,z^1)}  \right)^{-1}
\int\limits_0^1                               dz^1
\exp{{\scriptstyle {1 \over 2}} \phi(0,z^1)} \cos \pi mz^1
\label{dm}  \\
f_{3n} & = & \left( \int\limits_0^t         dz^0
\exp{{\scriptstyle {1 \over 2}} \phi(z^0,1)}  \right)^{-1}
\int\limits_0^t                               dz^0
\exp{{\scriptstyle {1 \over 2}} \phi(z^0,1)}
\cos \frac{\pi n z^0 }{t}  \nonumber \\
f_{4m} & = &
\left( \int\limits_0^1  dz^1
\exp{{\scriptstyle {1 \over 2}} \phi(t,z^1)}  \right)^{-1}
\int\limits_0^1                               dz^1
\exp{{\scriptstyle {1 \over 2}} \phi(t,z^1)} \cos \pi mz^1 \nonumber
\end{eqnarray}
Expanding the Liouville field $\phi$ in modes on $\widehat{M}_t$
$$
\phi = \phi_0 + \frac{2}{\sqrt{t}} {\sum\limits_{m,n \geq 0}}'
\phi_{nm} \cos \frac{\pi nz^0}{t} \cos \pi mz^1
$$
one finally has
\begin{eqnarray}
W[c_{RT}] & = & \int\limits_{0}^{\infty} dt \;\,\eta (t)^{1- {d \over 2}}
\int\limits_{- \infty}^{ + \infty} \sqrt{t}\; d\phi_0
\int {\prod_{n,m \geq 0}}' d\phi_{nm}    \nonumber \\
& & \times \exp \left(
-{25 - d \over 96 \pi}\left(
2\!\!\!\!\!\sum_{\mbox{\scriptsize $m$ or $n=0$}}
+{\sum_{n,m\geq 1}}\right)\left[ \left({n^2\pi^2\over t^2}+
m^2 \pi^2 \right) {\phi_{nm}}^2 \right]  \right) \nonumber \\
& & \times \exp \left( - {1 \over 4 \pi \alpha'}
S[ \widehat{g}_t, \widetilde{x}_i[\phi] ] \right)
\label{dn} \\
& & \times \exp \left( {7 + d \over 4 \sqrt{t} }
{\sum_{\mbox{ \scriptsize $n,m$ even}}}' \phi_{nm}
 +{7 +d \over 8}\phi_0 \right)  \nonumber \\
& & \times \exp \left(-\mu {\rm e}^{\phi_0}\int d^2z\exp\left(
\frac{2}{\sqrt{t}} {\sum\limits_{m,n \geq 0}}'
\phi_{nm} \cos \frac{\pi nz^0}{t} \cos \pi mz^1\right)\right) \nonumber
\end{eqnarray}
where $S[ \widehat{g}_t, \widetilde{x}_i[\phi] ]$ is given by the equations
(\ref{dl}),(\ref{dm}).

Since the boundary conditions (\ref{dbc}) are invariant with respect
to Weyl transformations with a conformal factor constant along boundary,
the zero mode $\phi_0$
enters the effective action only through the
corner anomaly and the Liouville interaction terms.
In both cases $\mu=0$, $\mu > 0$ the integration over $\phi_0$ can
be performed exactly.

For $\mu=0$, $\phi_0$ couples only via the corner anomaly term which
is independent of the Teichm\"{u}ller
parameter and of the parameters of $c_{RT}$.
Hence the only contribution
of $\phi_0$ is the term $\sqrt{t}$ coming from the $t$-dependence of
the zero mode integration measure and
some infinite constant which can be absorbed into
an overall normalization factor. Up to this constant one has
\begin{equation}
W[c_{RT}]
 =  \int\limits_{0}^{\infty} dt \;\,\eta (t)^{1- {d \over 2}}
\sqrt{t} Z(R,T;t)\;\;\;,
\label{wco}
\end{equation}
where
\begin{equation}
 Z(R,T;t) =
 \int {\cal D}'\phi \exp
\left( - {b \over 2\pi^2}\int d^2z (\nabla \phi)^2
       + \frac{c}{4} \sum\limits_{\mbox{\scriptsize corners}} \phi(z_i)
       - {1 \over \epsilon^2} V[\widetilde{\phi}]
       \right)\;\;\;,
\label{sa}
\end{equation}
In the formula above, prime in the symbol of measure means that the
integration is over all fields orthogonal to the zero mode subspace and
\begin{eqnarray}
b & = & \frac{(25-d)\pi}{48}\;\;\;, \nonumber \\
c& = & \frac{7 +d}{8}\;\;\;, \nonumber \\
V[\widetilde{\phi}] & = & \frac{1}{2R^2} S[\widehat{g}_t,\widetilde{x}_i[\phi]
]
\;\;\;,\;\; \widetilde{\phi} = \phi_{| \partial M} \nonumber\;\;.
\end{eqnarray}

In the presence of the exponential Liouville term in the action
($\mu >0$), the
integration over $\phi_0$ can be performed by means of the standard
techniques \cite{liouv,dhoker}. Using the formula

 \begin{equation}
\int_{-\infty}^{\infty}d\phi_0\ \exp\left(c\phi_0-\alpha{\rm e}^
{\phi_0}\right) = \Gamma(c)\ {\rm e}^{-c\ln\alpha}
\end{equation}
one gets
\begin{equation}
W[c_{RT}]
 = \mu^{-c} \Gamma(c)
  \int\limits_{0}^{\infty} dt \;\,\eta (t)^{1- {d \over 2}}
  t^{-c} Z(R,T;t)\;\;\;,
  \label{wcp}
\end{equation}
\begin{eqnarray}
Z(R,T;t) =
\int {\cal D}'\phi & & \!\!\!\!\!
\exp
\left( - {b \over 2\pi^2}\int d^2z (\nabla \phi)^2
       + \frac{c}{4} \sum\limits_{\mbox{\scriptsize corners}} \phi(z_i)
       - {1 \over \epsilon^2} V[\widetilde{\phi}]
       \right) \nonumber \\
& \times & \exp \left(
        - c \ln \left( {1\over t} \int d^2 z \exp(\phi) \right) \right) \;\;.
\label{saa}
\end{eqnarray}

Despite a considerable simplification in the $\phi$-dependence of
$V[\widetilde{\phi}] $ due to the linear
structure of boundary conditions (\ref{dj}), the "potential" part
of the effective action is extremely complicated. Actually it
is nonanalytical in terms of modes $\phi_{nm}$. However one can
easily show by explicit calculations that the functional
$V[\widetilde{\phi}] $ has
a true minimum at $\phi_{nm} \equiv 0$ which makes  semiclassical
calculations possible.

\section{Semiclassical Approximation for $\mu=0$}

The special feature of the Polyakov model making the semiclassical
calculations nontrivial is the universality of the coefficient in
front of the Liouville action. Since it is a dimensionless constant
uniquely determined by the number of target space dimensions, the only
available "loop" expansion in this sector corresponds to the formal
limit $d \rightarrow  -\infty$
\cite{dop}. On  the other hand the peculiar form
of the boundary conditions (\ref{dk}),(\ref{di})
and the decoupling of the zero mode from the $x$-sector lead to
the conclusion that the conformal factor should be regarded as an
infinite dimensional "Teichm\"{u}ller parameter" rather than a
physical degree of freedom. With this interpretation an
appropriate expansion parameter for a physically meaningful semiclassical
approximation is $\epsilon \equiv \frac{\sqrt{2\pi\alpha'}}{R}$.

In this section we analyse in more detail the semiclassical approximation
in the case $\mu =0$.
The corresponding consideration for the nonvanishing Liouville interaction
are presented in Sect.5.
In the  case of rectangular loop the "potential" term has an absolute minimum
$\widetilde{\phi}_{cl} \equiv 0$ which coincides with the boundary value
of the minimum of kinetic term. One can expect that in the limit
$\epsilon \rightarrow 0$
the main contribution to the functional integral (\ref{sa}) comes from
fluctuations around the classical configuration $\phi \equiv 0$.
Inserting the formal expansion
$$
V[\widetilde{\phi}] = V[\widetilde{\phi}_{cl}] +
\sum\limits_{n \geq 2} {1 \over n!} V^{(n)}[\widetilde{\phi}]
$$
into (\ref{sa}) and changing variables $\phi \rightarrow \epsilon \phi$
one gets
\begin{eqnarray}
Z(R,T;t)
 & = & \exp \left(- {1 \over \epsilon^2}  V[\widetilde{\phi}_{cl}]
\right)  \label{sc} \\
& \times &
\int {\cal D}'\phi \exp
\left( - {\epsilon^2 b \over 2\pi^2}\int d^2z (\nabla \phi)^2
       + \frac{\epsilon c}{4} \sum\limits_{\mbox{\scriptsize corners}}
          \phi(z_i)
       - {1 \over 2} V^{(2)}[\widetilde{\phi}] \right) \nonumber \\
& &\ \  \times \  \exp \left(
       - {\epsilon \over 3!} V^{(3)}[\widetilde{\phi}]
       - {\epsilon^2 \over 4!} V^{(4)}[\widetilde{\phi}] - ...
       \right)\;\;\;. \nonumber
\end{eqnarray}

The standard reasoning would lead to the conclusion that the 1-loop
approximation is given by dropping all terms with positive powers
of $\epsilon$. The resulting gaussian integral is, however, strongly
divergent and requires some regularization procedure. Our method
to deal with this problem consists in retaining at the 1-loop level
all terms quadratic and linear in $\phi$. Thus, the 1-loop approximation
is given by
\begin{eqnarray}
Z_{\mbox{\scriptsize 1-loop}}(R,T;t) & = &
\exp \left(- {1 \over \epsilon^2}  V[\widetilde{\phi}_{cl}] \right)
\label{sd} \\
&\times &
\int {\cal D}'\phi \exp
\left( - {\epsilon^2 b \over 2\pi^2}\int d^2z (\nabla \phi)^2
       + \frac{\epsilon c}{4} \sum\limits_{\mbox{\scriptsize corners}}
        \phi(z_i)
       - {1 \over 2} V^{(2)}[\widetilde{\phi}] \right)\;\;\;.
       \nonumber
\end{eqnarray}
The higher loop corrections are then defined by appropriate Wick contractions
of the interaction terms in (\ref{sc})
calculated in the gaussian model (\ref{sd}).
This prescription slightly differs from the usual loop expansion.
In fact, due to the presence of
higher order terms in (\ref{sd}) a fixed $n$-loop
correction may contribute to several terms in the $\epsilon$-expansion.
Moreover this additional $\epsilon$-dependence may
destroy the power expansion we would like to develop. The basic
consistency condition of the loop expansion given by (\ref{sd}) requires
that contributions from higher corrections lead to a formal
expansion of the following form:
\begin{equation}
Z(R,T;t) = \exp \left( -{1 \over \epsilon^2}
\left[ V[\widetilde{\phi_{cl}}] + \sum\limits_{n \geq 1} \epsilon^n
f_n(\epsilon) \right] \right)\;\;\;,
\label{se}
\end{equation}
where $f_n(\epsilon)$ have at most logarithmic dependence on $\epsilon$
for all $n\geq 1$. When this condition is satisfied the semiclassical
approximation is given by
\begin{equation}
Z_{\mbox{\scriptsize s.c.}}(R,T;t)
= \exp \left[ -{1 \over \epsilon^2}
\left( V[\widetilde{\phi_{cl}}] + \epsilon
f_1(\epsilon) \right) \right]\;\;\;.
\label{sf}
\end{equation}

In the rest of this section we shall analyse the 1-loop correction
in the limit of long rectangular loop. The interpretation of our
result as a semiclassical approximation is based on the
assumption that the condition (\ref{se}) is satisfied and that the only
contribution to the coefficient $f_1(\epsilon)$ comes from the 1-loop
correction (\ref{sd}). The explicit calculations presented below suggest
that in the limit ${T \over R} \rightarrow \infty$ this assumption is well
justified. An exact proof, however, requires a systematic analysis of
higher order corrections and is beyond the scope of the present
paper.

Expanding the functional (\ref{dl}) to the second power in the Liouville field
around $\phi_{cl} \equiv 0$ and using explicit formula (\ref{dn}) one gets
\begin{equation}
Z_{\mbox{\scriptsize 1-loop}}(R,T;t) =
e^{ - {1 \over 2 \epsilon^2} \left( {T^2 \over R^2 t} + t \right) }
 \int {\prod_{n,m \geq 0}}' d\phi_{nm}
e^{ - S_{\mbox{\scriptsize 1-loop}}[\phi]}      \;\;\;,
\label{sg}
\end{equation}
where
\begin{eqnarray}
S_{\mbox{\scriptsize 1-loop}}[\phi] & = &
{\epsilon^2 b \over 2}\left(
2\!\!\!\sum_{\mbox{\scriptsize $m$ or $n=0$}}
+ {\sum_{n,m \geq 1}}\right)\left[ \left( {n^2 \over t^2}
+m^2  \right) {\phi_{nm}}^2 \right] \nonumber \\
&+& {2\epsilon c \over  \sqrt{t} }
{\sum_{\mbox{ \scriptsize $n,m$ even}}}' \phi_{nm}   \nonumber \\
&+&
{T^2\over 2tR^2}\sum_{n=1}^{\infty}
\left[{ \tanh{\pi n\over 2t}\over \pi n}\left(
\sum_{\mbox{\scriptsize $m$ even}} \phi_{nm}\right)^2+
 {{\rm coth}{\pi n\over 2t}\over \pi n}
 \left(\sum_{\mbox{\scriptsize $m$ odd}} \phi_{nm}\right)^2\right]
 \label{sh} \\
&+& {1 \over 2 t}\sum_{m=1}^{\infty}
\left[{ \tanh{\pi mt\over 2}\over \pi m}\left(
\sum_{\mbox{\scriptsize $n$ even}} \phi_{nm}\right)^2+
{{\rm coth}{\pi mt\over 2}\over \pi m}
\left(\sum_{\mbox{\scriptsize $n$ odd}} \phi_{nm}\right)^2
\right] \;\; . \nonumber
\end{eqnarray}

It will be shown in the Appendix A that the conformal anomaly (the second
 factor in equation (\ref{sh}) does not contribute to the result in
the limit ${T \over R}\to\infty$ and that the last sum in  (\ref{sh})
 can also be neglected. Hence, in the case of long
 rectangular loop the 1-loop effective action can be replaced in
 the formula (\ref{sg}) by
\begin{eqnarray}
{S'}_{\mbox{\scriptsize 1-loop}}[\phi] & = &
{\epsilon^2 b \over 2}\left(
2\!\!\!\sum_{\mbox{\scriptsize $m$ or $n=0$}}
+ {\sum_{n,m \geq 1}}\right)\left[ \left( {n^2 \over t^2}
+m^2  \right) {\phi_{nm}}^2 \right]
\label{smg}  \\
&+&
{T^2\over 2tR^2}\sum_{n=1}^{\infty}
\left[{ \tanh{\pi n\over 2t}\over \pi n}\left(
\sum_{\mbox{\scriptsize $m$ even}} \phi_{nm}\right)^2+
 {{\rm coth}{\pi n\over 2t}\over \pi n}
 \left(\sum_{\mbox{\scriptsize $m$ odd}} \phi_{nm}\right)^2\right] \;\;\;.
 \nonumber
\end{eqnarray}

The resulting  integral can be decomposed into an infinite product
of gaussian integrals corresponding to each fixed value of the index $n$.
The integration over the modes $\left\{ \phi_{0,m}\right\}_{m \geq 1}$
yields just a constant factor. For every $n\geq 1$ the integration over
$\left\{ \phi_{n,m}\right\}_{m \geq 0}$ can be done by means of the
formula
\begin{equation}
{\rm det} \left(\matrix{\alpha_1+\beta&\beta&\ldots&\beta\cr
                  \beta&\alpha_2+\beta&\ldots&\beta\cr
                  \ldots&\ldots&\ldots&\ldots\cr
                  \beta&\beta&\ldots&\alpha_n+\beta\cr}\right)
= \left(1+\sum_j{\beta\over \alpha_j}\right)\prod_{i=1}^n \alpha_i \;\;\;.
\label{det}
\label{si}
\end{equation}
Performing all integrations one gets
\begin{equation}
             Z_{\mbox{\scriptsize 1-loop}}(R,T;t) \approx
e^{ - {1 \over 2 \epsilon^2} \left( {T^2 \over R^2 t} + t \right)}\times
      I\times I_o \times I_e \;\;\;,
\label{sj}
\end{equation}
where the infinite products $I, I_e, I_o$ are given by
\begin{eqnarray*}
I &=& \left[{\prod_{n,m \geq 0}}'\,\,\frac{\epsilon^2b}{2}
        \left({n^2 \over t^2} +m^2 \right)\right]^{-{1\over 2}}
        \;=\; t^{-{1\over 2}} \eta(t)^{-{1\over 2}}\;\;,\nonumber \\
I_e& = &        \prod_{n\geq 1}\left(1+
{ T^2\tanh {\pi n\over 2t}\over \epsilon^2bR^2\pi n t}
\left(\frac{t^2}{2n^2}+\sum_{\mbox{\scriptsize $m$ even $\ge 2$}}
        {1 \over {{n^2 \over t^2} +m^2 }}\right)\right)^{-{1\over 2}}\;\;,\\
I_o &  = &       \prod_{n\geq 1} \left(1+
{ T^2\coth {\pi n\over 2t}\over \epsilon^2bR^2\pi n t}
\sum_{\mbox{\scriptsize $m$ odd}}
        {1 \over {{n^2 \over t^2} +m^2}}\right)^{-{1\over 2}} \;\;\;. \nonumber
\label{deta}
\end{eqnarray*}
Using
$$\frac{1}{2a^2}+\sum_{k_{even}\ge 2}\frac{1}{k^2+a^2} =
\frac{\pi}{4a} {\rm coth} (\pi a/2)$$
\begin{equation}
\sum_{k_{odd}> 0}\frac{1}{k^2+a^2} =\frac{\pi}{4a} \tanh (\pi a/2)
\label{infsums}
\end{equation}
we find
\begin{eqnarray}
\ln I_o & = &
-\frac{1}{2}\sum_n\ln\left(1+\frac{T^2}{4\epsilon^2 b R^2  n^2}\right) \;\;,
\label{slo}\\
\ln I_e & =  &
-\frac{1}{2}\sum_n\ln\left(1+\frac{T^2}{ 4\epsilon^2 b R^2 n^2}\right)\ .
\label{sle}
\end{eqnarray}
Using the exact formula (which can be proven by differentiating
the both sides over $\alpha$)
\begin{equation}
\sum_{n\ge 1}\ln\left(1+\frac{\alpha^2}{n^2}\right)= \pi\alpha
-\ln 2\pi\alpha
+\ln\left(1-e^{-2\pi\alpha}\right)
\label{for}
\end{equation}
 one gets the exact result for the series (\ref{slo}) and (\ref{sle}).
Hence we can write
\begin{equation}
\ln  I_o = \ln I_e = -{\pi T\over 4\sqrt{b}\epsilon R}
 + O\left(\ln {\pi T\over \sqrt{b}\epsilon R}\right)
\label{sloa}
\end{equation}

Collecting all the results one gets in the
limit ${T\over R}\to \infty$
\begin{equation}
 Z_{\mbox{\scriptsize 1-loop}}(R,T;t)
 \approx  t^{-{1\over 2}} \eta(t)^{-{1\over 2}}
\exp \left[ - {1 \over  \epsilon^2}
\left(
 {T^2 \over 2R^2 t} + {t\over 2}
 + \epsilon f_1^{\mbox{\scriptsize 1-loop}} +
\epsilon^2 f_2^{\mbox{\scriptsize 1-loop}}
 \right)
 \right]\;,
\label{sm}
\end{equation}
where
\begin{eqnarray*}
f_1^{\mbox{\scriptsize 1-loop}}(\epsilon)& =&
{2\pi T \over 4\sqrt{b}R} \ ,\\
f_2^{\mbox{\scriptsize 1-loop}}(\epsilon)& =&
O\left(\ln \left({T \over \epsilon R}\right) \right)\ .
\end{eqnarray*}
The results above show that the consistency condition (\ref{se})
is satisfied on the 1-loop level. Thus, under the assumption
$f_1(\epsilon) = f_1^{\mbox{\scriptsize 1-loop}}(\epsilon)$,
the semiclassical approximation is given by dropping in the
formula (\ref{sm}) all terms proportional to $\epsilon^2$.
Note that these terms do  not contribute to the static potential anyhow.

 Inserting (\ref{sm}) into (\ref{wco}),
 changing variables $t \rightarrow \tau = {t\over T}$ and neglecting all
 $O\left(\ln \left({T \over \sqrt{2\pi\alpha'}}\right) \right)$-terms in
 the exponent one gets
\begin{equation}
W[c_{RT}] \approx  \int\limits_{0}^{\infty}
d\tau \; e^{-TG(R,\tau)}\;\;\;,
\label{sn}
\end{equation}
where
$$
G(R,\tau) = {1\over 4\pi\alpha'\tau} + \left(
{R^2\over 4\pi\alpha'} - {(d-1)\pi\over 24} \right) \tau +
{2\pi \over 4\sqrt{2\pi\alpha'b}}\ .
$$
For $R$ above the critical distance
\begin{equation}
R_c = \sqrt{{\pi(d-1)2\pi\alpha' \over 12}}\ ,
\label{prom}
\end{equation}
the equation
$$
-{1\over 4\pi\alpha'\tau^2} +
{R^2\over 4\pi\alpha'} - {(d-1)\pi\over 24} =0\ ,
$$
has a solution
$$
\tau_{\mbox{\scriptsize min}}(R)=\left(R^2-R^2_c\right)^{-1/2}
$$
 which is an absolute minimum of the function $G(R,\tau)$.
It follows that the large $T$ asymptotic of the integral (\ref{sn}) is
given by the saddle point approximation
\begin{equation}
W[c_{RT}]\ \Tinfty\ {\rm e}^{-TV(R)}
\label{sr}
\end{equation}
where
\begin{equation}
V(R)=\frac{1}{2\pi\alpha'}\sqrt{R^2-R^2_c}+\frac{\pi}{2\sqrt{2\pi\alpha' b}}
\label{pote}
\end{equation}
According to \cite{w} $V(R)$ can be interpreted as the static quark potential
in the Polyakov string approximation of QCD.
The special feature of the Polyakov
string is the appearance of a  nonvanishing perimeter term.

\section{Semiclassical Approximation for $\mu>0$}

The approximation scheme developed in the previous section can be also
applied to the functional integral (\ref{saa}). As before one can
expand around $\phi_{nm} \equiv 0$. In the present case this is not
a minimum of the whole action, but since the Liouville term is bounded
from below the main contribution to the integral (\ref{saa}) comes
from a small neighbourhood of the minimum of the "potential" term.
Thus, expanding the logarithm and the exponential function one gets
\begin{equation}
Z_{\mbox{\scriptsize 1-loop}}(R,T;t) \approx
e^{ - {1 \over 2 \epsilon^2} \left( {T^2 \over R^2 t} + t \right) }
 \int {\prod_{n,m \geq 0}}' d\phi_{nm}
e^{ - S_{\mbox{\scriptsize 1-loop}}[\phi]}      \;\;\;,
\label{s2g}
\end{equation}
where
\begin{eqnarray*}
S_{\mbox{\scriptsize 1-loop}}[\phi] & = &
{\epsilon^2 b \over 2}\left(
2\!\!\!\sum_{\mbox{\scriptsize $m$ or $n=0$}}
+ {\sum_{n,m \geq 1}}
\right)\left[ \left( {n^2 \over t^2}
+m^2 +\frac{c}{bt} \right) {\phi_{nm}}^2 \right] \nonumber \\
&+& {2\epsilon c \over  \sqrt{t} }
{\sum_{\mbox{ \scriptsize $n,m$ even}}}' \phi_{nm}   \nonumber \\
&+&
{T^2\over 2tR^2}\sum_{n=1}^{\infty}
\left[{ \tanh{\pi n\over 2t}\over \pi n}\left(
\sum_{\mbox{\scriptsize $m$ even}} \phi_{nm}\right)^2+
 {{\rm coth}{\pi n\over 2t}\over \pi n}
 \left(\sum_{\mbox{\scriptsize $m$ odd}} \phi_{nm}\right)^2\right] \\
&+& {1 \over 2 t}\sum_{m=1}^{\infty}
\left[{ \tanh{\pi mt\over 2}\over \pi m}\left(
\sum_{\mbox{\scriptsize $n$ even}} \phi_{nm}\right)^2+
{{\rm coth}{\pi mt\over 2}\over \pi m}
\left(\sum_{\mbox{\scriptsize $n$ odd}} \phi_{nm}\right)^2
\right] \;\; . \nonumber
\end{eqnarray*}

As before the conformal anomaly (the second
 factor in the equation above does not contribute to the result in
the limit ${T \over R}\to\infty$ and the last sum
 can also be neglected. Hence, for large ${T \over R}$ the 1-loop
 effective action takes the form
\begin{eqnarray}
{S'}_{\mbox{\scriptsize 1-loop}}[\phi] & = &
{\epsilon^2 b \over 2}\left(
2\!\!\!\sum_{\mbox{\scriptsize $m$ or $n=0$}}
+ {\sum_{n,m \geq 1}}
\right)\left[ \left( {n^2\pi^2 \over t^2}
    +m^2 +\frac{c}{bt} \right) {\phi_{nm}}^2\right]
\label{smgpp}  \\
&+&
{T^2\over 2tR^2}\sum_{n=1}^{\infty}
\left[{ \tanh{\pi n\over 2t}\over \pi n}\left(
\sum_{\mbox{\scriptsize $m$ even}} \phi_{nm}\right)^2+
 {{\rm coth}{\pi n\over 2t}\over \pi n}
 \left(\sum_{\mbox{\scriptsize $m$ odd}} \phi_{nm}\right)^2\right] \;\;\;.
 \nonumber
\end{eqnarray}

Using the result (\ref{si}) and the  formula
$$
\det(\Delta+M^2)=\det(\Delta){\rm e}^{\ln\delta\times M^2\int d^2z\sqrt{g}}
$$
we get (up to a constant)
\begin{equation}
Z_{\mbox{\scriptsize 1-loop}}(R,T;t) \approx
e^{ - {1 \over 2 \epsilon^2} \left( {T^2 \over R^2 t} + t \right)}\times
      I\times I_o \times I_e \;\;\;,
\label{s2j}
\end{equation}
where
\begin{eqnarray*}
   I& = &  t^{-{1\over 2}} \eta(t)^{-{1\over 2}}\ \nonumber \\
\ln I_o & = &
-\frac{1}{2}\sum_n\ln\left(1+\frac{T^2\coth\frac{\pi n}{2t}}
{4\epsilon^2 b
R^2  nt} \frac{\tanh\frac{\pi}{2}\sqrt{\frac{n^2}{t^2}+\frac{c}{bt}}}
{\sqrt{\frac{n^2}{t^2}+\frac{c}{bt}}}
\right) \;\;,
\label{s2lo}\\
\ln I_e & =  &
-\frac{1}{2}\sum_n\ln\left(1+\frac{T^2\tanh\frac{\pi n}{2t}}
{4\epsilon^2 b
R^2  nt} \frac{\coth\frac{\pi}{2}\sqrt{\frac{n^2}{t^2}+\frac{c}{bt}}}
{\sqrt{\frac{n^2}{t^2}+\frac{c}{bt}}}
\right) \nonumber
\end{eqnarray*}
The series above are convergent for all positive values of
parameters ${T \over \epsilon R}, t$.
Summing over $n$ from 1 to $\sqrt{t}$ and from $\sqrt{t}$ to
$\infty$ and using the formula (\ref{for})
one gets for ${t \over T}  = \mbox{const}$, $T\rightarrow \infty$
\begin{equation}
\ln  I_o = \ln I_e = -{\pi T\over 4\sqrt{b}\epsilon R}
 + O\left(\sqrt{T} \ln {\pi T\over \sqrt{b}\epsilon R}\right)\;\;\;.
\label{wyzn}
\end{equation}
Inserting (\ref{wyzn}) into the formula (\ref{s2j}) one obtains
$$
 Z_{\mbox{\scriptsize 1-loop}}(R,T;t)
 \approx  t^{-{1\over 2}} \eta(t)^{-{1\over 2}}
\exp \left[ - {1 \over  \epsilon^2}
\left(
 {T^2 \over 2R^2 t} + {t\over 2}
 + \epsilon {2\pi T \over 4\sqrt{b} R} +
\epsilon^2 O\left( \sqrt{T} \ln {T\over \epsilon R} \right)
 \right)
 \right]\;.
$$
The formula above exactly coincides with the corresponding one (\ref{sm})
derived in the case $\mu = 0$. Thus repeating all the calculations
of the previous section one arrives at the result (\ref{pote}).

\section{Conclusions}

The comparison of our results with the predictions of
the string models from L\"{u}scher's universality class \cite{luscher}
shows two differences. The first one consists in the presence of
the nonvanishing "universal"
perimeter term. The second more important feature is that the longitudinal
mode of noncritical Polyakov string manifests itself  simply as an
additional degree of freedom. This can be easily seen comparing the
formulae (\ref{prom},\ref{pote}) with the result
found in the Nambu-Goto string model by
Alvarez \cite{alv} and Arvis \cite{arvis}.
$$
R^{\mbox{\scriptsize N-G}}_c = \sqrt{{\pi(d-2)2\pi\alpha' \over 12}}\;\;,
$$
$$
V^{\mbox{\scriptsize N-G}}(R)
=\frac{1}{2\pi\alpha'}\sqrt{R^2-{R^{\mbox{\scriptsize N-G}}_c}^2}\;\;\;.
$$
Our semiclassical result leads to the conclusion that the Polyakov sum
over bordered surfaces does not provide a good ansatz for the Wilson
loop. This is especially
striking in two dimensions where the formula (\ref{pote}) essentially differs
from the linear static potential predicted by QCD.

In  string theory the static potential $V(R)$ is interpreted as the
ground state energy of the string with ends fixed at the distance $R$.
In the semiclassical approximation the expressions for the
rectangular Wilson loops (now interpreted as a string propagator) are
different for $\mu = 0$ and $\mu > 0$. Nevertheless, the large $T$
asymptotics  exactly coincide. A closer analysis shows that not
only the ground state energies but the whole spectra are identical.
This result is consistent with the conjecture
\cite{dhoker} that the Liouville theory may be a complicated off-shell
extension of a basically simple model.

The investigations of the static potential presented in this paper leave
the open problem of higher loop corrections.  A more detailed analysis of
this point is in fact necessary for a full justification of the assumption made
in our calculations. It is also interesting to understand the structure of
the nonstandard loop expansion introduced in Section 4. The 1-loop
calculations suggest that it is quite regular. In fact the only
renormalization we need in the gaussian models (\ref{smg}),
(\ref{smgpp}) generating the
loop expansions concerns the "bulk" part of the determinant in the
formulae (\ref{sj}) and (\ref{s2j}) respectively..

The question arises how general the methods employed in this paper are.
The perturbation method introduced in the beginning of Sect.4
is based on the assumption
that for each value of the Teichm\"{u}ller parameter $t \in R_+$,
 there exists a minimum of the functional
$S_{\mbox{\scriptsize eff}}[c,e^{\phi}\widehat{g}_t]$ (\ref{dh}).
In the general case of a model
surface $M$ with a boundary $\partial M$ mapped to a given contour $c
\subset R^d$ by $\widetilde{c}:\partial M \rightarrow c$
the corresponding functional
$S[\widetilde{c},\widetilde{\phi};t]$ can be regarded as a functional
$S[\widetilde{c},\gamma ;t]$ on the space $\mbox{Diff}(\partial M) \times
T_M$ where $T_M$ is the Teichm\"{u}ller space of the surface $M$. Note
that $S[\widetilde{c},\gamma ;t]$ is just the energy functional from the
theory of harmonic maps
\cite{j} evaluated at the solution $x_{cl}$ of the
Dirichlet boundary problem $\Delta_t x_{cl} = 0 , {x_{cl}}_{|\partial M} =
\widetilde{c}\circ \gamma$. The minimal parametrization
$\gamma_{\mbox{\scriptsize min}}(t)$ is then uniquely determined by the
solution of the free boundary problem for the contour $c$. Sufficient
conditions for the existence of solution to this problem
are usually expressed in terms of some regularity conditions for the
contour $c$
\cite{j}. Now, the existence of a minimum of the function
$S[\widetilde{c},\gamma_{\mbox{\scriptsize min}}(t);t]$ on $T_M$ is
equivalent to the existence of a solution of the Plateau problem (with
the fixed topological type)
\cite{j}. The resulting conditions for the contour $c$
precisely coincide with that one encounters in the semiclassical
calculations in the Schild-Eguchi string model \cite{lsw}.
The existence of a minimum of the interaction term provides merely a
justification for the perturbation scheme developed in Sect.4. For
concrete calculations one needs an explicit solution of the free
boundary problem for each point in the Teichm\"{u}ller space.
This is the main technical limitation of
the semiclassical calculations in
string models. Nevertheless, using the methods developed in the
case of long rectangular loop one can address all the questions of
the program initiated in Ref.\cite{lsw}.

As was mentioned in the introduction, semiclassical calculations of the
static potential can be used as the simplest test for any definition
of the sum over surfaces. In the case of the conceptually simple
but complicated theory formulated in
Sect.2 this test yields surprisingly good results. This leads us
to the main conclusion of this paper that in the range $1 <d<25$
the model derived in Sect.2 is a good candidate for
a well defined sum over surfaces in the continuum formulation.
It seems that the program of semiclassical calculations sketched
above can provide some additional information about
the quantum theory of conformal factor and in particular
the relevance of the Liouville interaction.
However, the most
intriguing question is whether the sum over surfaces yields a
consistent quantum theory of noncritical string. The work on
this problem is in progress and will be presented elsewhere.

\section*{Acknowledgements}

One of us (Z.J.) is grateful to Professor A.Verjovsky for enlightening
discussions on the variational problems related to this work.
We   would
like to thank Professor Abdus Salam, the International Atomic Energy Agency and
UNESCO for  support during our stay at the International Centre of Theoretical
Physics where this work was carried out.

\section*{Appendix A}

In this Appendix we will prove that the corner anomaly
 does not give any contribution to the quark potential
in the limit $R/T\to 0$.

Let us consider the even-even sector (the only one where the corner
enters).
\begin{eqnarray*}
 S_{\mbox{\scriptsize ee}}&=&
\frac{b}{2}
\left(
2\!\!\!\sum_{\mbox{\scriptsize $m$ or $n=0$}}
+ {\sum_{n,m \geq 2}}\right)_{\mbox{\scriptsize $m$,$n$ even}}
\left[ \left( {n^2 \over t^2}
+m^2  \right) {\phi_{nm}}^2 \right]
\\
&+&{T^2\over \apd\pi t}
\sum_{n\, \mbox{\scriptsize even}\, \ge 2}
\frac{ \tanh{\pi n\over 2t}}{n}
\left( \sum_{m\ \mbox{\scriptsize even}} \phi_{nm} \right)^2+
 \frac{c}{\sqrt{t}}\sum_{mn\ \mbox{\scriptsize even}} \phi_{nm}\;\;\;.
\end{eqnarray*}

We use the well known formula for the Gaussian functional integration:
$$
\int\prod_i\ d\phi_i\ \exp\ (-\phi_iA_{ij}\phi_j+b_i\phi_i)=
{\rm const\ (det}A)^{-1/2}\ \exp(b_iA_{ij}^{-1}b_j/4)       \;\;.
$$
The determinant has been considered in the main text so we will take into
account only the argument of the exponential. As before we can treat
separately each n and sum over it at the end. We have to remember that
$m$ and $n$ run over even integers only.

It is easy to prove
$$
\frac{1}{4} b_iA_{ij}^{-1}b_j=\frac{c^2}{4t}(1\ 1\ \ldots\ 1)
{\pmatrix{\alpha_1+\beta&\beta &\ldots &\beta\cr
         \beta&\alpha_2+\beta &\ldots &\beta\cr
         \ldots&\ldots&\ldots&\ldots\cr
         \beta&\beta&\ldots &\alpha_n+\beta\cr}}^{-1}
\pmatrix{1\cr 1\cr \ldots\cr 1} $$
$$
 =\frac{c^2}{4t}
\frac{\sum \alpha_m^{-1}}{1+\beta\sum \alpha_m^{-1}}
$$
where
$$
\alpha_m=\frac{b}{2}\left(\frac{n^2}{t^2}+m^2\right)\ \ ,
\alpha_0=\frac{bn^2}{t^2}
$$
$$
\beta=\frac{T^2\tanh\frac{\pi n}{2t}}{\apd \pi tn}\;\;\;.
$$

Using the equation (\ref{infsums}) we get
the total result (we changed $n$ into $n/2$ so that now
$n$ runs over all positive integers)
$$
-\frac{c^2 }{16 bt}\sum_{n=1}^{\infty}\frac{\frac{\pi n}{t}\coth
\frac{\pi n}{t}{\rm e}^{-\delta n^2/t^2}}
{\frac{n^2}{t^2}+\frac{T^2}{32\pi\alpha' bt^2}}
$$
We have introduced the regularization by the kinetic term - after removing
the regulator the result in the limit $T\to\infty$
, $t/T=$const is just a function of $t/T$ and not $T\times$function$(t/T)$ so
the whole contribution
from  the anomaly on the corners to the potential can be neglected.

At the end we would like to justify the fact that in the limit $R/T\to 0$
we dropped the infinite sum proportional to $R^2$ in eq. (\ref{sh}).
The structure of the eq. (\ref{sh}) can be presented in the following form:
\begin{eqnarray*}
A/T^2& =&A_0+\frac{R^2}{T^2}\delta A=
\pmatrix{A_{11mm'} &0&0&\ldots\cr
         0&A_{22mm'}&0&\ldots\ldots\cr
        \ldots&\ldots&\ldots&\ldots\cr}\\
& +&\frac{R^2}{T^2}\pmatrix{0&0&A_{13mm'}&0&A_{15mm'}&\ldots\cr
             0&0&0&A_{24mm'}&0&\ldots\cr
             A_{31mm'}&0&0&0&A_{35mm'}&\ldots\cr
             \ldots&\ldots&\ldots&\ldots&\ldots&\ldots\cr}
\end{eqnarray*}
where each entry with given values of $nn'$ corresponds to an infinite
matrix in $mm'$. We expand in $R^2$ the determinant of $A$:
$$
\ln\det A\approx \ln\det A_0+\frac{R^2}{T^2}{\rm Tr}\left(A_0^{-1}
\delta A\right) $$

The first term was considered in the main text while the second
(which is anyway proportional to $R^2/T^2$) vanishes since $\delta A$
is off-diagonal. It means that we can safely drop the whole $R^2$
correction to the determinant of $A$.

\end{document}